# Graph Compression Using Pattern Matching Techniques

Rushabh Jitendrakumar Shah
Department of Computer Science
California State University Dominguez Hills
Carson, CA, USA
rshah13@toromail.csudh.edu

*Abstract:* Graphs can be used to represent a wide variety of data belonging to different domains. Graphs can capture the relationship among data in an efficient way, and have been widely used. In recent times, with the advent of Big Data, there has been a need to store and compute on large data sets efficiently. However, considering the size of the data sets in question, finding optimal methods to store and process the data has been a challenge. Therefore, in this paper, we study different graph compression techniques and propose novel algorithms to do the same. Specifically, given a graph G = (V, E), where V is the set of vertices and E is the set of edges, and |V| = n, we propose techniques to compress the adjacency matrix representation of the graph. Our algorithms are based on finding patterns within the adjacency matrix data, and replacing the common patterns with specific markers. All the techniques proposed here are lossless compression of graphs. Based on the experimental results, it is observed that our proposed techniques achieve almost 70% compression as compared to adjacency matrix representation. The results show that large graphs can be efficiently stored in smaller memory and exploit the parallel processing power of compute nodes as well as efficiently transfer data between resources.

## I. Introduction

In the current era of Big Data and Internet of Things, there has been an exponential increase in the number of devices being connected together and the amount of data being generated and exchanged. With the huge size of the data, there are related challenges to storing and managing the data efficiently for effective data mining and computation. There are various choices of data structures to store the data being made available through different resources. Using graphs to represent such data is one of the choices available.

Graphs have been extensively used in different domains to represent both simple and complex data. With graphs being able to model and represent a plethora of information, it has become an essential data structure for a number of applications. Therefore, focus on different graph algorithms have been in the forefront of problem solving techniques. With the advent of non-conventional compute devices, including Graphics Processing Units i.e., GPUs, novel approaches have been taken to design parallel and distributed algorithms to solve the graphs problems [9] [10]. Therefore, using graphs as a data structure has become common in the field of Computer Science. Graphs can be stored using both basic and advanced data structures. Using arrays, or linked lists, or a combination of both are commonly used methods to store graphs.

The data being stored in graphs in recent times can be characterized as Big Data. Hence, it takes huge amount of space to store the graph data. Computing devices in recent times have improved from research that has contributed to increasing the capacities of the same. This has made it easier to store larger data on faster memory. However, with the advent of Big Data, the challenge of fitting data on compute nodes has come to the forefront of computer science research again. To extract knowledge and extract information from the Big Data sets, the data has to be stored on compute nodes and analyzed [15] [23]. However, due to the size of the data sets and the current standards in computer hardware this has become a challenge [12] [16]. This is profound in the process of utilizing multicore devices, specifically GPUs, where the available memory on the compute nodes is many orders of magnitude smaller than what is available on a daily-use computer [7] [13] [14]. Hence, storing graphs efficiently is of utmost importance. Therefore, in this paper we focus on techniques to compress graphs without losing any information of the relationship between the entities of the data sets. Specifically, we identify different patterns that exist within the graph data, and then reduce the space requirements by replacing the larger patterns with a specific marker that represents the original data. The graph can be uncompressed by using the reverse method for compression. In addition, the compressed data itself can also be used to perform analysis and calculations on the graph as well.

The outline of the paper is as follows. In Section II, the related work is included; this focusses on previous research in the area of graph compression. The proposed techniques are included in Section III. The experimental results and related analysis are provided in Section IV. Conclusion and future work are given in Section V.

## II. Related work

Graphs represent data from various domains and improving the storage and retrieval of data from graphs have a significant effect on a wide variety of algorithms [17] [18] [19]. Hence, graphs have been studied for a long time and there exists research techniques the focus on graph compression among others. The graph compression techniques also take into account the structure of the graph, which can vary significantly depending on the domain under consideration.

In this paper we consider graphs that represent real world data. One of the main properties of such graphs are that they are sparse in nature. Therefore, as an example, the research is focused on graphs that represent online social networks or the web i.e., graphs with power law distribution properties.

Exhibiting the Web as a graph with addresses as nodes and hyperlinks as edges among nodes is common. Usually, a significant amount of compression can be achieved by modifying adjacency lists using pointers to other similar lists; the differences i.e., missing and added information is also stored in addition to the common data [5]. There are also techniques that emphasize on using a web graph compression scheme that focused on finding nodes with related sets of neighbors [1].

Another technique is based on using common neighborhoods and exploiting locality information to compress graphs with power law distribution. The method leverages the fact that many pages on a common host have related sets of neighbors [4].

Properties of web graphs can also be exploited in techniques based on lexicographical ordering [3]. It can be observed that proximal pages in URL lexicographical ordering often have similar neighborhoods. This lexicographical locality property allowed the use of gap encodings when compressing edges. In order to further improve compression, new orderings can be developed that combine host information and Gray/lexicographic orderings [3].

Using techniques based on lexicographic ordering along with neighborhood information is effective in compressing graphs [24]; however, since most real world data do not have any natural order, this mechanism does not work for the graphs being considered in this paper.

Quadtree based technique to compress graph data has been used before [6] [8] [12]. However, the topological information of the graph has not been taken into consideration. In this paper, by exploiting the structural patterns in the graphs, we introduce algorithms that lead to better graph compression.

### III. PROPOSED TECHNIQUES

Graphs can be represented using various representations. Each of the data structures has its advantages and disadvantages with respect to the amount of memory required to store the data and the ease of access to the data [21] [22]. Depending on the requirements, sometimes it is relevant to store the data in larger data structures that requires more space but provides efficient access to the data at the time of computation. On the other hand, if memory is limited, then using data structures that provide storage using the least amount of memory is desired. Using less memory in turn can require complicated mechanisms to access the data, but with computation becoming cheaper with the advancements of devices, this issue is less significant for certain cases.

Although there are different data structures representing graphs in various ways, in this Section we consider a few of the common data structures that capture the essence of most of the other variations. We then introduce our idea of using a modification of the same to reduce the required memory for storage. In general, graphs can be represented using one of the following three data structures:

1. Edge Lists
2. Adjacency List
3. Adjacency Matrix

Edge List is a simple way to denote a graph, and most real-world graphs are represented using this method. Given a graph $G = (V, E)$, where $\{V\}$ is the set of vertices and $\{E\}$ is the set of edges, it is a list of all the edges of the graph under consideration in the form of $(V_i, V_j)$, where $V_i, V_j \in \{V\}$. The number of such entries in the list would be $|E|$.

Adjacency list is group of unordered list used to denote a finite graph. Each part of the list defines the set of neighbors in the graph. In the simplest form, there is a list for each vertex $V_i \in \{V\}$, which consists of a set of vertices $\{V_x\}$, where there is a direct edge between $V_i$ and $V_j$, where $V_j \in \{V_x\}$.

Adjacency matrix is the most commonly used representation for storing graphs. The adjacency matrix is a square matrix, of size n x n, where $|V| = n$. Each of the elements in the adjacency matrix represents whether pairs of the vertices are adjacent or not in the graph being represented.

In this paper, we focus on using variations of the adjacency matrix. The adjacency matrix representation is straight forward and provides constant time access to adjacency information between a pair of vertices. From the time complexity perspective, the performance of adjacency matrix as a data structure is efficient. It takes $O(1)$ time to query for an edge in the graph. However, with the advantage of providing efficient access to the topological data, using an adjacency matrix might not be suitable for all graphs belonging to different domains. Depending on the structure and characteristics of the graphs being represented, adjacency matrix might be an efficient choice or might add too much overhead in terms of memory usage.

$$\begin{pmatrix}
1 & 1 & 1 & 1 & 0 & 0 & 0 & 0 & 0 & 0 & 0 & 0 & 0 & 0 & 0 \\
0 & 0 & 0 & 0 & 1 & 1 & 1 & 1 & 0 & 0 & 0 & 0 & 0 & 0 & 0 \\
0 & 0 & 0 & 0 & 0 & 0 & 0 & 1 & 1 & 1 & 1 & 0 & 0 & 0 & 0 \\
0 & 0 & 0 & 0 & 0 & 0 & 0 & 0 & 0 & 0 & 0 & 1 & 1 & 1 & 1 \\
1 & 0 & 0 & 0 & 1 & 0 & 0 & 0 & 1 & 0 & 0 & 0 & 1 & 0 & 0 \\
0 & 1 & 0 & 0 & 0 & 1 & 0 & 0 & 0 & 1 & 0 & 0 & 0 & 1 & 0 \\
0 & 0 & 1 & 0 & 0 & 0 & 1 & 0 & 0 & 0 & 1 & 0 & 0 & 0 & 1 \\
0 & 0 & 0 & 1 & 0 & 0 & 0 & 1 & 0 & 0 & 0 & 1 & 0 & 0 & 0 & 1 \\
1 & 0 & 0 & 0 & 0 & 1 & 0 & 0 & 0 & 0 & 1 & 0 & 0 & 0 & 0 & 1 \\
0 & 1 & 0 & 0 & 0 & 0 & 1 & 0 & 0 & 0 & 0 & 1 & 1 & 0 & 0 & 0 \\
0 & 0 & 1 & 0 & 0 & 0 & 0 & 1 & 0 & 1 & 0 & 0 & 0 & 0 & 1 & 0 \\
0 & 0 & 0 & 1 & 1 & 0 & 0 & 0 & 1 & 0 & 0 & 0 & 0 & 1 & 0 & 0
\end{pmatrix}$$

Fig. 1: Sample Adjacency Matrix Representation

An example of an adjacency matrix representation of a graph is given in Fig. 1. The 1 values indicate existence of an edge between the vertices given by the row and column numbers of the entry; a value of 0 indicates the absence of any edge between the same. As it can be observed, for the given graph, the adjacency matrix contains 25% 1 values and the rest 75% 0 values. A graph where the density of the edges as compared to the maximum possible edges is comparatively low, is considered sparse. Sparse graphs usually have more 0's than 1's in the adjacency matrix representation. Therefore, using adjacency matrix for sparse graphs might have a higher memory overhead for storing values that represent elements that do not exist. Most real world graphs are sparse. In the example shown in Fig. 1, there are still 25% of the maximum number of edges present in the graph, and by definition this is sparse. In actual practice, most real world data are sparse and the density of edges are typically less than 5% and mostly in the 1% to 2% range. Hence, while this adjacency representation provides efficient edge information, this data structure may not be suitable for storing large graphs in limited memory compute nodes or devices.

In this Section we introduce our idea of exploiting topological pattern similarities to reduce the memory requirement overhead. The core principle focuses on identifying "patterns" that are sequence of adjacency information, and replacing all the patterns with markers that have a lower space requirement. This idea is developed for sparse matrices, but would work for any graph in general. For denser graphs, the different number of patterns that can potentially match would be a large number, thereby reducing the effect. In any case, since the research focuses on representations and analysis on real world data sets, the graphs representing such data are sparse. Therefore, our technique works well for such domain data.

Algorithm 1: Matching patterns in adjacency matrix
Input: Adjacency matrix A[ ] for G = (V, E), Pattern Set {P}
Output: Compressed matrix $A_c$[ ] for G = (V, E)
For each row in A[ ]
    Divide row into chunks, Size(chunk)=Size(patterns)
    For each chunk in the row
        If matchPattern(chunk, $P_i$), $P_i \in$ {P}.
            Insert 1
            Insert Indicator_$P_i$
        Else
            Insert 0
            Insert Chunk

Algorithm 1 splits each of the rows into chunks of sizes equal to those of the patterns, and are matched with patterns in the set of patterns given by {P}. Each successful match results in replacing the chunk with an indicator of the pattern and a leading 1 indicating the next bits belonging to an indicator. In the case of where the chunk does not match with any of the patterns in {P}, the raw data is stored preceding with a 0, which indicates the presence of raw data.

The resulting compression of the graph mainly depends on the number of chunks in the adjacency matrix that matches with the pre-defined patterns. In addition, for all matches and non-matches, an additional 1-bit data, either a 0 or a 1 is added to indicate the succeeding pattern indicator or raw data. Therefore, for all the non-matches, the performance of the compression technique is degraded. The final dominating factor in the calculation of the compressed matrix size is the number of bits required to represent the patterns that are matched. In case of patterns of size n bits, the indicator would require a size of $\log_2 n$ bits. For our analysis, we consider the patterns to be of size 8, 16, 32 and 64 bits. The respective number of bits for each of the patterns in the pattern indicator would be 3, 4, 5 and 6. Depending on the size of the graphs under consideration, the size of the patterns and the pattern indicators would vary. In this paper, we consider real world data, and therefore, for experimental purposes, we chose graphs of size 1024, 2048, 4096 and 8192. The graphs can be generated either using a graph generator suite or locally following certain properties [2]. Hence, we choose the pattern sizes to be either 32 and 64. Therefore, the pattern indicator sizes are either 5 bits or 6 bits.

Also, depending on the size of the patterns, the number of patterns to be matched varies. The potential number of comparisons is given by (Size of the matrix / Size of the pattern). So for a graph of 1024 nodes and pattern size 8 we have to do 131,072 comparisons; for pattern size of 32 we have to do 32,768 comparisons. To reduce the overhead of the number of patterns to be matched, we select patterns of size 32 in this paper.

Choosing the patterns should reflect the topology of the graph being considered. If random patterns are considered, then finding matches for the patterns would be difficult and the corresponding compression would be inefficient. In this paper we consider real world graph, which are sparse in nature. Therefore, the patterns also represent the data under consideration.

In this paper we consider 3 different types of patterns, each of size 32. Pattern 1 consists of a two 1's in a 32-bit combination. The leading bit is always a 1, and there is another 1 at another position. Hence, we have 31 patterns using such a method. Since the graphs being considered are sparse, we also include a pattern with all 32 bits as 0's. Hence, in total there are 32 patterns for the first set. For pattern 2, we consider combinations with just a single 1 bit and the remaining all 0's. Since a 1 bit can be placed at the 32 different positions in the 32-bit pattern, there are 32 patterns in the pattern 2 set. Finally, in the pattern 3, we combine the patterns of 1 and 2. This set includes combinations that have all 0's, the only single 1's and finally two 1's. The total number of patterns in this set is 64.

Therefore, the sizes of the pattern indicators for patterns 1 and 2 is 5, as it requires 5 bits to represent 32 different combinations. For the pattern 3, the pattern indicator size if 6 bits since there are 64 different combinations in this set.

IV. EXPERIMENTAL RESULTS

Using the proposed algorithm, graphs of different sizes are compressed; the results of the compression and observations are discussed in this Section.

We consider graphs with 1024, 2048, 4096 and 8192 nodes. Although, real world graphs have much higher number of nodes, previous research has shown the connected components of most real world graphs are no larger than the sizes considered here

[10] [11]. Therefore, the results of these compression techniques apply to larger data sets as well [20].

The experimental results using pattern 1 are shown in Fig, 2. For all the different sizes of the graphs, about 21% compression was achieved. Since the effect of the compression methods depend on the number of patterns that could be matched, we report the values of the patterns found. The values are plotted against the maximum possible patterns that are present, as shown in Fig. 2. On an average, for the 1024 node graph, ~9500 patterns were found and ~23500 not found. For the other graphs of sizes 2048, 4096 and 8192, the values of found and not found were respectively ~38000 & ~92000, ~155000 & ~370000, and ~620000 & ~1470000, respectively. Using the pattern 1, overall about 21% compression was achieved as compared to the adjacency matrix size.

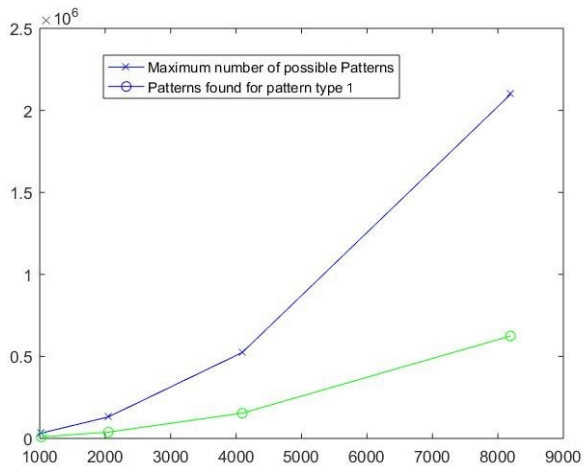

Fig. 2: Matchings using Pattern 1

The compression algorithm is also implemented using the pattern 2 set as described before. For pattern 2 we were able to achieve greater than 45% of compression for all the 4 different size graphs under consideration.

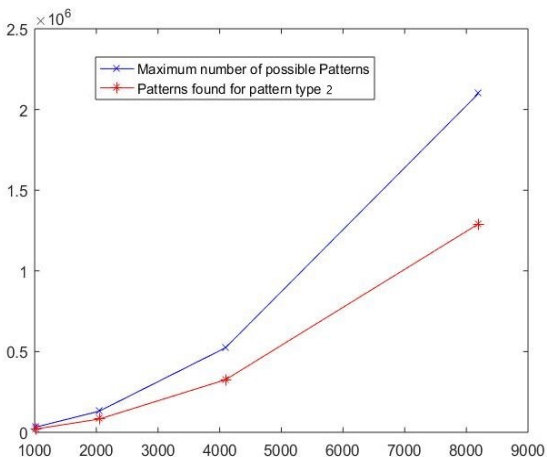

Fig. 3: Matchings using Pattern 2

The results for the number of patterns that were matched against the total number of patterns present for this case is depicted in Fig. 3. It is evident that the difference between the potential matches and actual matches in this case is less than those while considering pattern 1.

Finally, using the pattern 3 combinations, which are the union of patterns 1 and 2, an estimated 70% compression was achieved using the proposed algorithm. The results of the implementation are provided in Fig. 4. It can be observed that the difference between the number of patterns present and those actually matched has decreased significantly.

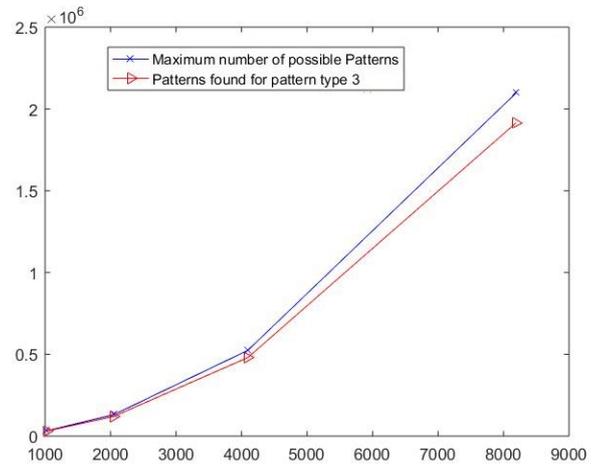

Fig. 4: Matchings using Pattern 3

Fig. 5 presents the 3 different patterns together. It is evident from the graph that the performance of pattern 3 is better and it produces the highest number of matches with the data chunks and also produces the most efficient compression.

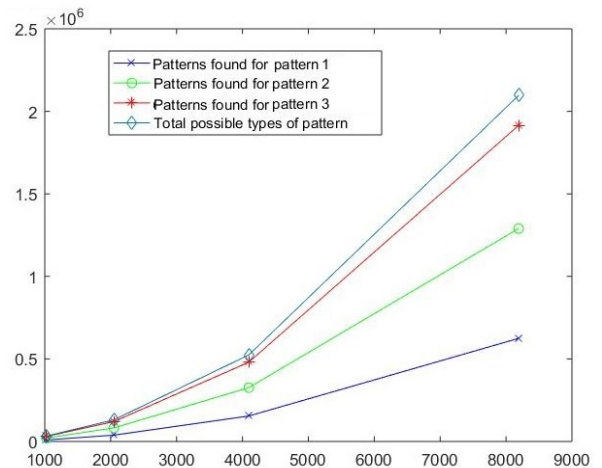

Fig. 5: Comparison of the matchings for different patterns

V. CONCLUSION

Graphs have been in the forefront of data structures that are used to represent a plethora of varied information. In the era of Big Data, storing and analyzing graphs is essential to generate and gather intelligence and information from data. Therefore, storing graphs has been a major challenge with non-conventional compute devices. In this paper we provide an algorithm to compress graphs based on the adjacency matrix representation of the same. The core idea involves finding patterns and replacing the same with identifies in order to reduce the memory requirement. We consider sparse graphs, as those represent real world data, and our algorithm implementation achieves up to 70% compression as compared to the adjacency matrix representation of the graphs. Currently we are using a static set of patterns as the set to find and match patterns with; in the future, using a dynamic set that can change depending on the domain of data would be interesting to explore.